\documentclass[preprintnumbers,amsmath,amssymb,twocolumn,groupedaddress,floatfix]{revtex4}

\usepackage{graphicx}
\usepackage{bm}
\usepackage{slashed}
\usepackage{amsfonts}
\usepackage{graphicx,color}
\usepackage[colorlinks,linkcolor=blue,citecolor=green]{hyperref}
\usepackage{wrapfig}

\begin{document}

\title{Chiral phase transition in  linear sigma model with non-extensive statistical mechanics}
\author{Ke-Ming Shen}
\email{shenkm@mails.ccnu.edu.cn}
\affiliation{Key Laboratory of Quark $\&$ Lepton Physics (MOE) and Institute of Particle Physics,
Central China Normal University, Wuhan 430079, China}

\author{Hui Zhang}
\email{mr.zhanghui@mails.ccnu.edu.cn}
\affiliation{Key Laboratory of Quark $\&$ Lepton Physics (MOE) and Institute of Particle Physics,
Central China Normal University, Wuhan 430079, China}

\author{De-Fu Hou}
\affiliation{Key Laboratory of Quark $\&$ Lepton Physics (MOE) and Institute of Particle Physics,
Central China Normal University, Wuhan 430079, China}

\author{Ben-Wei Zhang}
\email{bwzhang@mail.ccnu.edu.cn}
\affiliation{Key Laboratory of Quark $\&$ Lepton Physics (MOE) and Institute of Particle Physics,
Central China Normal University, Wuhan 430079, China}

\author{En-Ke Wang}
\affiliation{Key Laboratory of Quark $\&$ Lepton Physics (MOE) and Institute of Particle Physics,
Central China Normal University, Wuhan 430079, China}

\date{\today}

\begin{abstract}

From the non-extensive statistical mechanics, we investigate the chiral phase transition at 
finite temperature $T$ and baryon chemical potential $\mu _B$ in the framework
of the linear sigma model. The corresponding non-extensive distribution,
based on Tsallis' statistics, is characterized by a dimensionless non-extensive
parameter, $q$, and the results  in the usual Boltzmann-Gibbs case are recovered when $q\to 1$.
The thermodynamics of the linear sigma model and its correspodning phase diagram are analysed.
At high temperature region, the critical temperature
$T_c$ is shown to decrease with increasing $q$ from
the phase diagram in the $(T,~\mu)$ plane.
However, larger values of $q$ causes the rise of $T_c$ at low temperature
but high chemical potential.
Moreover, it is found that $\mu$ different from zero corresponds to
a first-order phase transition while $\mu=0$ to a crossover one.
The critical endpoint (CEP) carries higher chemical potential
but lower temperature with $q$ increasing due to the non-extensive effects.
%
\end{abstract}

\maketitle

\section{Introduction}\label{sec:sec1}

~~Quantum Chromodynamics (QCD) is a basic theory of
describing the strong interactions among quarks and gluons, the fundamental
constituents of matter. 
More and more attentions have already been attracted to the QCD phase transition
both theoretically and experimentally
\cite{B-17,RV-16,L-15,ALICE-15,FLW-14,R-09,GM-05,SM-12}. 
Though experimental studies and lattice Monte-Carlo simulations have
made it possible to research on the phase diagram quantitatively,
there still remains uncertainty at high baryon density region \cite{P-09}.
Consequently, the phase transition is also
a vital topic in high energy physics, where the thermal vacuum created by heavy-ion collisions
differs from the one at zero temperature and chemical potential \cite{GM-05}. In order to
study certain essential features of it, the linear sigma model has been proposed to
illuminate the restoration of chiral symmetry and its spontaneous breaking \cite{SM-12}.

Near the phase transition boundary one should be cautious when using the
Boltzmann-Gibbs (BG) statistics for the appearance of critical
fluctuations due to a large correlation length \cite{SRS-98}.
It is of interest to investigate the phase transition in the formalism beyond conventional BG statistical mechanics. Recently a non-extensive statistics firstly proposed 
in Ref. \cite{T-88} has attracted a lot of attention and discussions~\cite{q-app}. In this formalism, instead of the exponential function, a generalized $q-$exponential function is defined as\cite{T-88,q-app},
\begin{eqnarray}
\exp_q(x):=[1+(1-q)x]^{\frac{1}{1-q}}~,
\label{q-exp}
\end{eqnarray}
where the parameter $q$ is called the non-extensive parameter, which accounts for
all possible factors violating assumptions of the usual BG case.
Its inverse function is also listed\cite{T-88,q-app},
\begin{eqnarray}
\ln_q(x):=\frac{x^{1-q}-1}{1-q}~.
\label{q-log}
\end{eqnarray}
Both of them return to the usual exponential and logarithm function with $q \to 1$.


The purpose of this paper is to clarify the non-extensive effects on physical quantities of
the chiral phase transition in the generalized linear sigma model. 
We focus on the situations where both of temperature and chemical potential are
not vanished, which then indicates the influence of the Tsallis distribution on the whole phase diagram
in the $(T,~\mu)$ plane. Whereas the non-extensive parameter $q$ is
still a phenomenological parameter \cite{q-app}, not only the case of $q>1$ but $q<1$, in
this work, are computed. For comparisons, we
have presented discussions
on this issue 
for finite temperature but vanishing chemical potential \cite{I-15}.
Given the consistency of non-extensive generalizations with the initially
BG approaches, we also list the results of $q=1$ which were investigated 
\cite{SMMR-01}. We close our researches with the comparisons
to the Non-extensive Nambu Jona-Lasinio Model ($q-$NJL model) \cite{RW-09} of the critical endpoint (CEP), whose location is still the
hot topic for experiments as well as its theoretical researches \cite{L-15,FLW-14}.

~~This paper is organized as follows. In Section~\ref{sec:sec2} we introduce the theoretical framework,
where the non-extensive $q-$linear sigma model is stated.
Their consequences for various thermodynamic
quantities with different non-extensive parameters, $q$, are explored in Section~\ref{sec:sec3};
more detailed discussions on the results are also contained. 
Section~\ref{sec:sec4}
is our brief summary and outlook.

\section{Theoretical Framework}\label{sec:sec2}

~~Within the linear sigma model, the chiral effective Lagrangian 
with quark degrees of freedom reads \cite{SMMR-01,BB-84,BB-85}
\begin{eqnarray}
\mathcal{L}&=&\bar{\psi}[i\gamma ^\mu \partial_\mu -g(\sigma+i\gamma_5 \vec{\tau}\cdot \vec{\pi})]\psi
\nonumber \\ &&+\frac{1}{2}(\partial_\mu \sigma \partial^\mu \sigma +
\partial_\mu \vec{\pi}\cdot \partial^\mu \vec{\pi})-U(\sigma, \vec{\pi}) ~,
\label{Lagrangian}
\end{eqnarray}
where $\psi =(u,d)$ stands for the spin$-\frac{1}{2}$ two flavors light quark fields,
the scalar field $\sigma$ and the pion field $\vec{\pi}=(\pi_1,\pi_2,\pi_3)$ together form
a chiral field $\Phi=(\sigma,\vec{\pi})$, with its potential
\begin{eqnarray}
U(\sigma, \vec{\pi})=\frac{\lambda^2}{4}(\sigma^2+\vec{\pi}^2-v^2)^2-H\sigma~.
\label{U-potential}
\end{eqnarray}
Considering the obvious symmetry breaking term $H\sigma=0$, $\cal{L}$ is invariant under chiral
$SU(2)_L\times SU(2)_R$ transformations. The chiral symmetry is spontaneously broken in the vacuum
with the expectation values: $\langle \sigma\rangle=f_\pi$ and $\langle \vec{\pi}\rangle=0$, where
$f_\pi=93$ MeV is the pion decay constant. By the partially conserved axial vector current (PCAC) relation\cite{SMMR-01} the 
quantity $v^2=f^2_\pi-m_{\pi}^2/\lambda^2$
with the constant $H=f_\pi m_\pi^2$ where $m_\pi=138$ MeV is the pion mass. The coupling
constant $\lambda^2$ is fixed as $20$ by $m_\sigma^2=2\lambda^2 f_\pi^2+m^2_\pi$ where
$m_\sigma=600$ MeV is the sigma mass. 
Another coupling constant $g$ is usually determined
by the requirement of the constituent quark mass in vacuum, $M_{vac}=gf_\pi$, which is about $1/3$ of the
nucleon mass, leading to $g\approx 3.3$. \cite{SMMR-01}

In order to investigate the temperature $T$ and the chemical potential
$\mu$ dependence in this model, let us consider a system of both quarks
and antiquarks in the thermodynamical equilibrium. Here quark chemical potential $\mu\equiv \mu_B/3$.
And the grand partition function goes like
\begin{eqnarray}
\mathcal{Z}&&=\textit{Tr} \exp [-\frac{\mathcal{H}-\mu \mathcal{N}}{T}] \nonumber\\
&&=\int \mathcal{D}\bar{\psi}\mathcal{D}\psi
\mathcal{D}\sigma\mathcal{D}\vec{\pi} \exp [\int_x (\mathcal{L}+\mu \bar{\psi}\gamma^0 \psi)] ~,
\label{Z-partition}
\end{eqnarray}
where $\int_x\equiv i\int_0^{1/T}dt\int_V d^3\vec{x}$ with $V$ is the volume of the system.

Thus the grand canonical potential can be obtained
\begin{equation}
\Omega (T, \mu) =-\frac{T}{V}\ln \mathcal{Z}=U(\sigma, \vec{\pi})+\Omega_{\bar{\psi}\psi}
\label{omega}
\end{equation}
with the (anti)quark contribution is
\begin{eqnarray}
\Omega_{\bar{\psi}\psi}(T, \mu)&=&-\nu \int \frac{d^3\vec{p}}{(2\pi)^3}
\{E+T\ln[1+e^{-(-\mu +E)/T}] \nonumber \\&&+T\ln[1+e^{-(\mu +E)/T}]\}
\label{Therm-potential}
\end{eqnarray}
where $\nu=12$ is the internal degrees of freedom of quarks 
and $E=\sqrt{p^2+M^2}$
is the valence (anti)quark energy, with the mass
of constituent (anti)quark defined as
\begin{eqnarray}
M^2=g^2(\sigma^2+\vec{\pi}^2)~.
\label{Mass}
\end{eqnarray} 
Here the first divergent 
term of $E$ is absorbed in the coupling constant in the results
which comes from the negative energy states of the Dirac sea.

It is inadequate to apply naively the BG statistics in such a system
critical fluctuations of energy and
particle numbers will appear as well as a large correlation.
In order to
investigate the phase transition of systems departing from the classical thermal equilibrium,
the non-extensive statistics 
\cite{T-88} is introduced. 
The so-called Tsallis entropy and density matrix are given, respectively, as
$S_q=k_BTr(\rho -\rho^q)/(q-1)$ and $\rho =\exp_q(-E/T)/Z_q$, where
$k_B$ is the Boltzmann constant (set to 1
for simplicity next), $q$ describes the degree of non-extensivity
and $Z_q$ is the corresponding generalized partition function.

Recently this generalized statistics has been of great interest theoretically 
\cite{LMM-98,B-01,NMT-11} and widely applied in many fields 
\cite{BP-05,PSA-07,LG-08,BM-12}. In the following we investigate the linear sigma model
within the non-extensive statistics. 
Firstly we re-write the partition function of Eq.(\ref{Z-partition}) as
\begin{eqnarray}
\mathcal{Z}_q&&=\textit{Tr} \exp_q [-\frac{\mathcal{H}-\mu \mathcal{N}}{T}] \nonumber\\
&&=\int \mathcal{D}\bar{\psi}\mathcal{D}\psi 
\mathcal{D}\sigma\mathcal{D}\vec{\pi} \exp_q [\int_x (\mathcal{L}+\mu \bar{\psi}\gamma^0 \psi)]
\label{Z-partition-q}
\end{eqnarray}
where the $q-$exponential is seen in Eq.(\ref{q-exp}). Considering
the $q-$thermodynamics \cite{Tsallis-book}
we have
\begin{eqnarray}
\Omega_{\bar{\psi}\psi}(T, \mu, q)&=&-\frac{T}{V}\ln_q\mathcal{Z}_q-U(\sigma, \vec{\pi}) \nonumber\\
&=&\sum _n \sum _p \{ \ln_q[\beta ^2(E_n^2+(E-\mu)^2)] \nonumber \\
&&+\ln_q[\beta^2(E_n^2+(E+\mu)^2)]\}~.
\label{Therm-potential-q-1}
\end{eqnarray}
Before carrying it out, we give out the generalized identities with respect to $q-$sums and integrals,
\begin{eqnarray}
\ln_q[\beta ^2(E_n^2+(E\pm\mu)^2)]&=&\int_1^{\beta^2(E\pm \mu)^2}
\frac{d \theta^2}{(\theta^2 +(2n+1)^2\pi^2)^q} \nonumber \\ &&+\ln_q[1+(2n+1)^2\pi^2]
\label{int-q}
\end{eqnarray}
and the generalized sum over $n$, in our assumptions, 
\begin{eqnarray}
\sum _n \frac{1}{(\theta^2 +(2n+1)^2\pi^2)^q}\approx \frac{1}{\theta}(\frac{1}{2}-
\frac{1}{(\exp_{2-q}(\theta)+1)^q})
\label{sum-n-q}
\end{eqnarray}
where $E_n=(2n+1)\pi T$ is used and the index $2-q$ appears because of the duality,
\begin{equation}
\exp_q(-x)\cdot\exp_{2-q}(x)=[1-(1-q)x]^{\frac{1}{1-q}}\cdot [1+(q-1)x]^{\frac{1}{q-1}}=1~.
\label{q-duality}
\end{equation}
Integrating over $\theta$ and dropping terms that are independent of $\beta$ and $\mu$,
we finally obtain
\begin{eqnarray}
\Omega_{\bar{\psi}\psi}(T, \mu, q)&=&-\nu \int \frac{d^3\vec{p}}{(2\pi)^3}
\{E+T\ln_q[1+\exp_q(-\frac{E-\mu}{T})] \nonumber \\ &&+T\ln_q[1+\exp_q(-\frac{E+\mu}{T})]\}~.
\label{Therm-potential-q}
\end{eqnarray}
 
In our calculations within the mean-field approximation, 
we follow 
Refs. \cite{CMM-96} and \cite{SMMR-01} where
the expectation value of the pion field is set to zero, $\vec{\pi}=0$.
By solving the gap equation,
\begin{equation}
\frac{\partial}{\partial \sigma}\Omega_{\bar{\psi}\psi}(T, \mu, q)|_{\sigma=\sigma_V}=0
\label{gap-equ}
\end{equation}
the value for constituent (anti)quark mass $M=g\sigma_V$ can be determined, which will be also
affected from different non-extensive parameters, $q$. 
Here we have replaced $\sigma$ and $\vec{\pi}$ in the exponent 
by their expectation values in the mean-field
approximation. 

With such a $q-$thermal effective potential, we then explore
the non-extensive effects on the physical quantities, as well as the phase transition,
in the linear sigma model. The numerical results will be shown in the next section.

\section{Results and Discussion}\label{sec:sec3}

~~In virtue of it there still exist fierce controversies over the possible
interpretations of the non-extensive parameter $q$, we shall discuss the non-extensive
effects in the $q-$linear sigma model for both the $q>1$ case and $q<1$. Meanwhile,
we give out the result of $q=1$ as the baseline for better understanding.

Worthy to note that the value of non-extensive parameter $q$ cannot
be much smaller than $1$ since in the expression of Eq.(\ref{Therm-potential-q}), whose corresponding
generalized exponential
\begin{equation}
\exp_q(-\frac{E+\mu}{T}):=[1-(1-q)\frac{(E+ \mu)}{T}]^{\frac{1}{1-q}}
\label{q-exp-1}
\end{equation}
where the part of the base should be larger than $0$: $1-(1-q)\frac{(E+ \mu)}{T}>0$, namely,
\begin{equation}
q>1-\frac{T}{(E+ \mu)}
\label{q-exp-2}
\end{equation}
Thus some upper limitation of energy of the integral in Eq.(\ref{Therm-potential}) should be given
in case of divergence when $q<1$. On the other hand, too much smaller values of $q$ are not
necessary to be computed physically during our investigation on the non-extensive effects
on the phase transition. Therefore, here we just list the results of $q=1.1,~1.05,~0.95$
and $q=1$ as baseline.

\begin{figure}[!htb]
	\includegraphics[width = 0.9\linewidth]{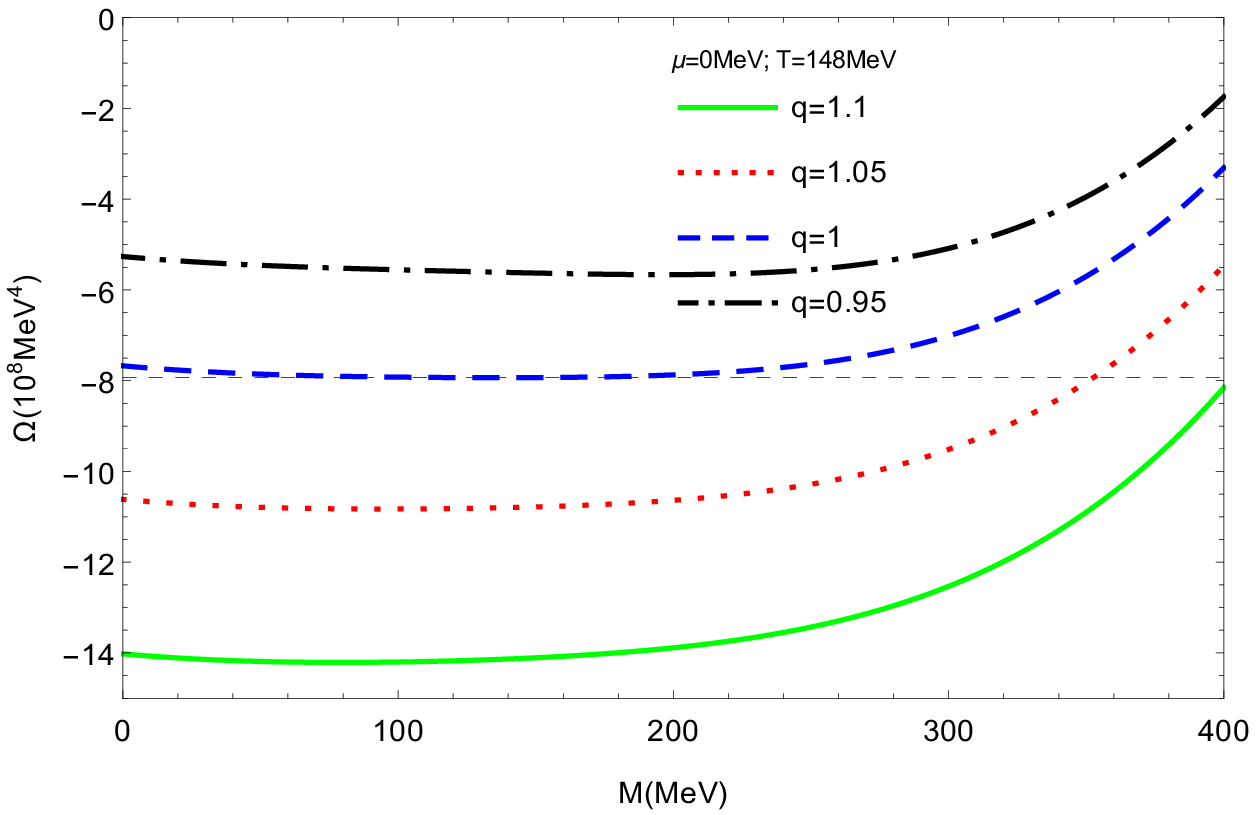}
	\includegraphics[width = 0.88\linewidth]{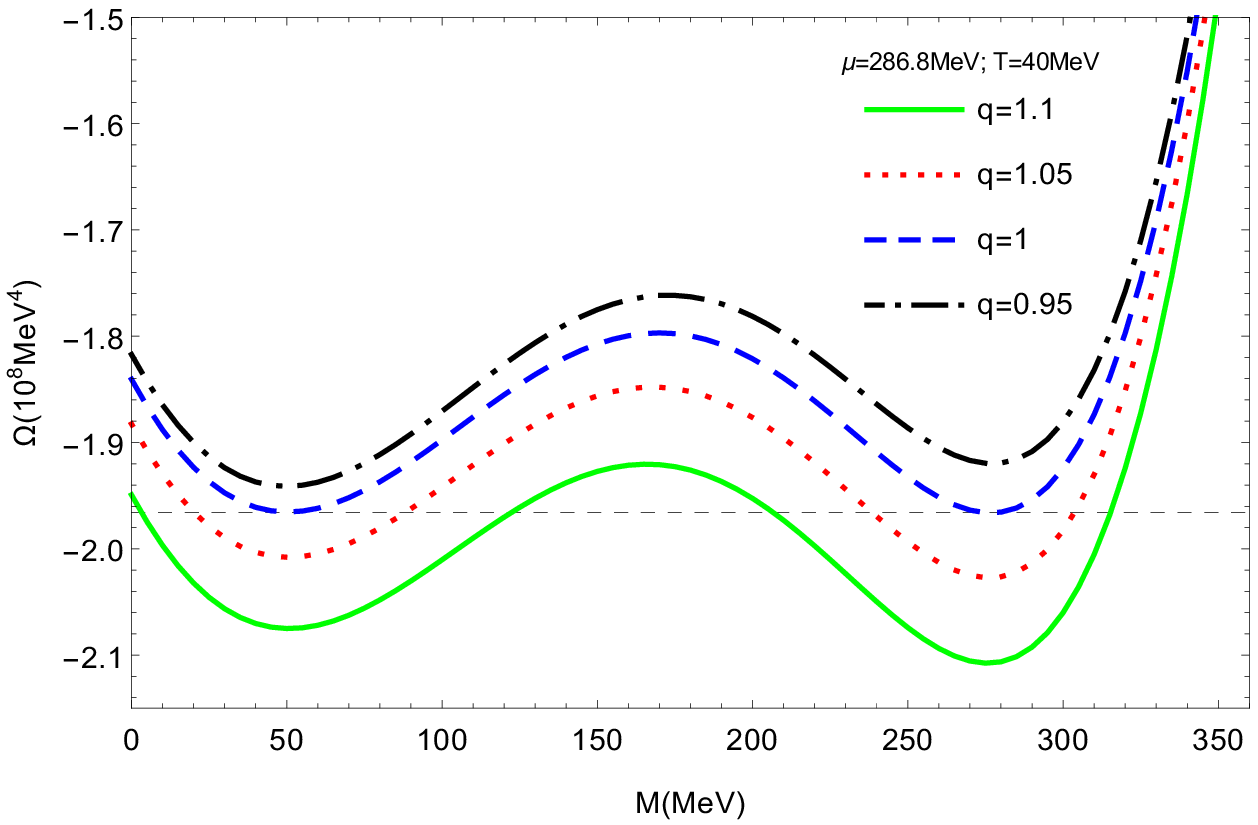}
\caption{The thermodynamical potentials $\Omega$, cf. Eq.(\ref{Therm-potential-q}) 
and (\ref{Therm-potential}), with respect to the
constituent (anti)quark mass $M$. Upper panel:
$T=148$ MeV and $\mu=0$. Lower panel:
$T=40$ MeV and $\mu=286.8$ MeV. We compare our results using
the parameters near the phase boundaries.
Both of them are analysed for $q=1.1$,
$q=1.05$, and $q=0.95$ with the case of $q=1$ as comparisons.}
\label{potential-M}
\end{figure}

We start our discussions with presenting in Fig.\ref{potential-M} the resulting
thermodynamical potential $\Omega$ as a function of $M$, the constituent (anti)quark
mass. Different $q$ evidently results in a large change of the thermodynamical potential
which shows the effects caused by non-extensivity are quite strong
whether the quark chemical potential vanishes or not. 

For the upper panel, the potentials with different $q$ are shown for
$T=148$ MeV and $\mu=0$. Locations of its minimum become smaller when $q$ gets larger.
This means that in the case of high temperature and low density,
correlations with the non-extensive $q-$version, shift the chiral condensation
toward smaller $M$. On the other hand (for the lower one), at low temperature but high density,
($T=40$ MeV and $\mu=286.8$ MeV),
the gap of potential between local (near $M=0$) and global (far from $M=0$) vacuum also increases as $q$ increases.
Worthy to note that, as seen in the lower panel, different $q$ nearly not affects
the position of global vacuum which should have nothing to do with the model itself.

\begin{figure}[!htb]
	\includegraphics[width = 0.88\linewidth]{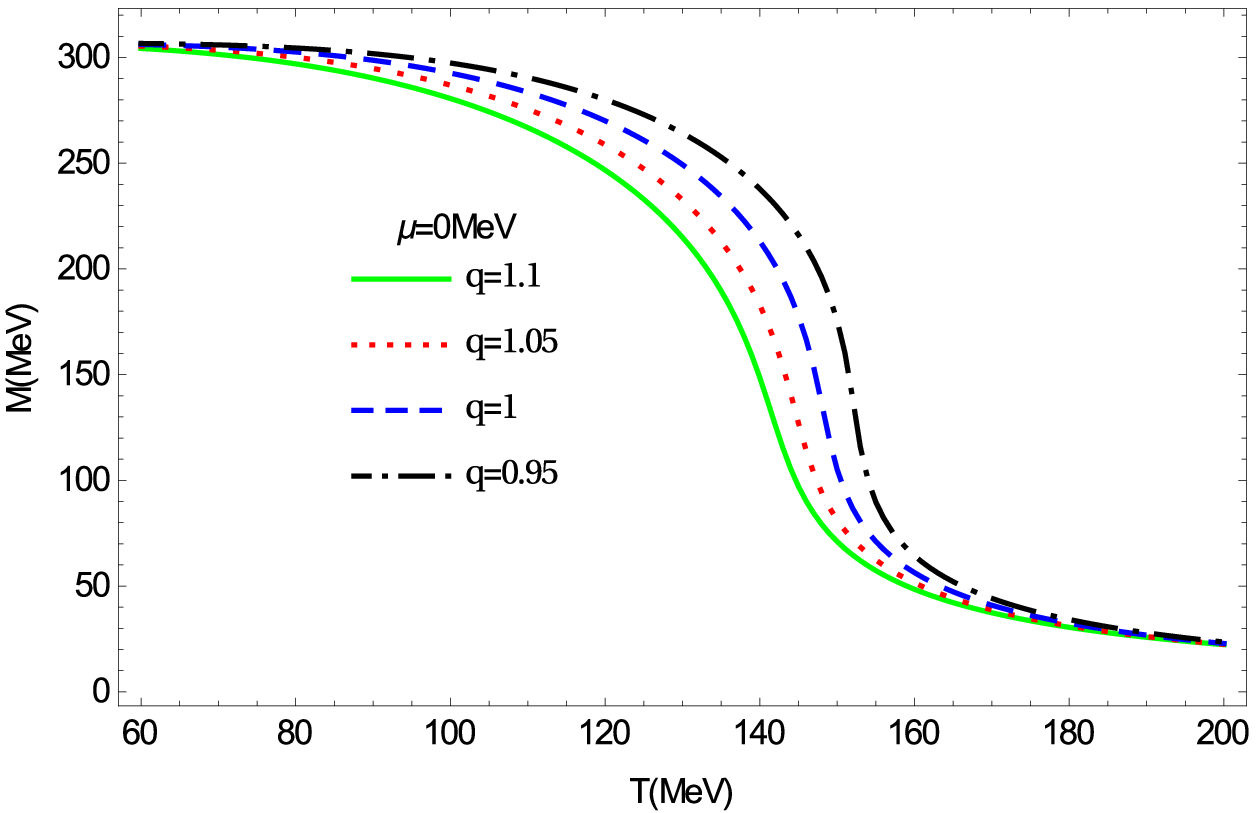}
	\includegraphics[width = 0.9\linewidth]{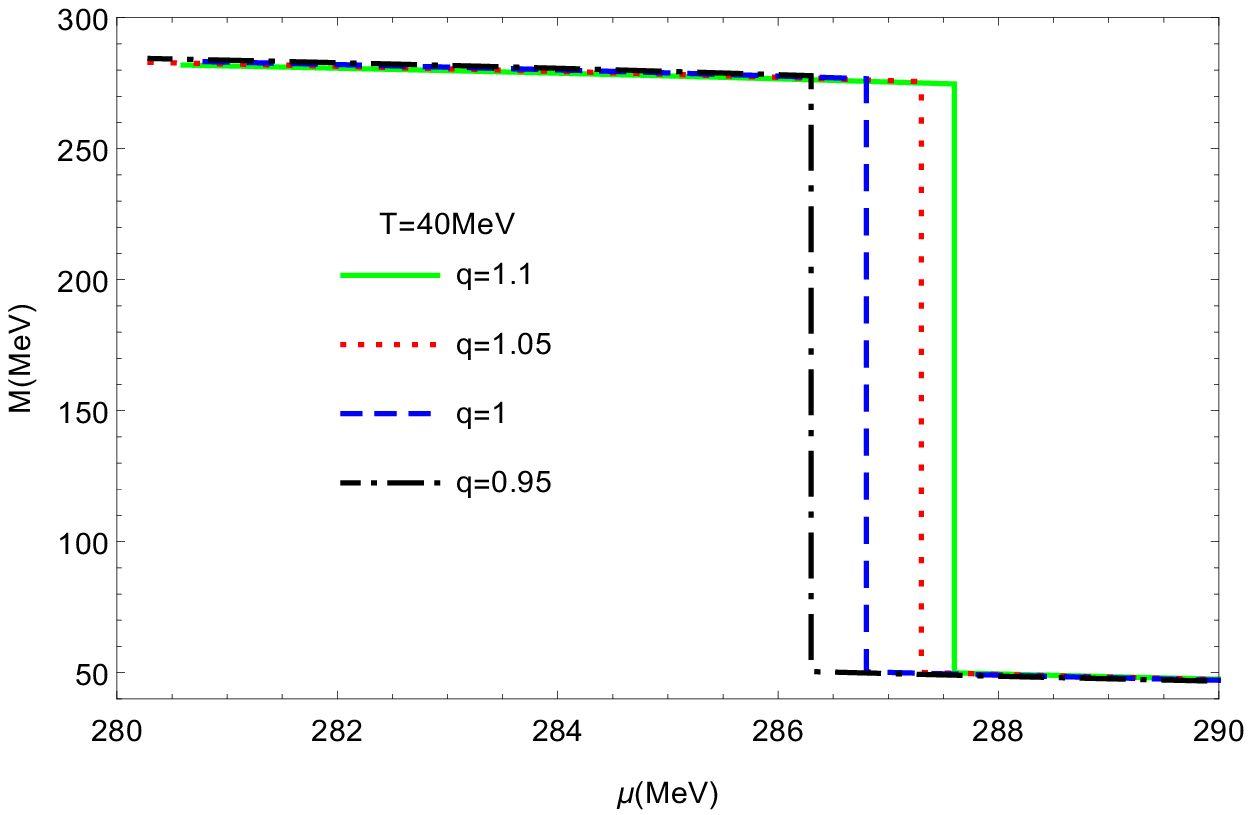}
\caption{The constituent (anti)quark mass $M$ as functions of the
temperature $T$ at $\mu=0$ (upper panel) and the chemical
potential $\mu$ at $T=40$ MeV (lower panel) for different $q$ as Fig.\ref{potential-M}.}
\label{M-effect}
\end{figure}

It is instructive to plot the $q-$effects on the constituent (anti)quark mass $M$ under
the temperature dependences as well as the chemical potential dependences,
which are clearly shown in Fig.\ref{M-effect}.
For the $T-$dependence ($\mu =0$), the values of $M$ change continuously with
the temperature $T$, which describes a typical crossover transition.
While for the $\mu-$dependence (where we set $T=40$ MeV), it shows a jump over the values of $M$, 
demonstrating a first-order phase transition.

The upper plot, at low density,
indicates the temperature dependence of $M$ for $q\neq 1$
is quite similar to the case of $q=1$, 
the usual BG situation.
Both the minimum and maximum of $M$ keep the same values
for different $q$. Nevertheless, the behaviour of all curves tells us
high temperature is required to restore the chiral symmetry for small $q$,
which agrees with the results 
of \cite{I-15}.

At the same time, for the low temperature case, the lower panel of Fig.\ref{M-effect} illustrates
the $\mu-$dependence of the constituent (anti)quark mass in the non-extensive linear sigma model 
for different non-extensive parameter $q$,
which is not done 
in Ref. \cite{I-15}. One easily observes an
analogous pattern characteristic to the above, while for the $\mu-$dependence,
increasing $q$ will also increase the value of phase transition chemical potential when $T=40$ MeV is fixed.
Moreover, for both of the cases, it is deserved to be mentioned
that only the system near the chiral phase transition
is well affected by non-extensive statistics.

In statistical physics, the critical properties of a thermodynamic system 
can be explored by studying the fluctuations of various observables.
Particularly, the fluctuations of the order parameter probe 
the order of the phase transition and the position of a possible critical end point.

\begin{figure}[!htb]
	\includegraphics[width = 0.9\linewidth]{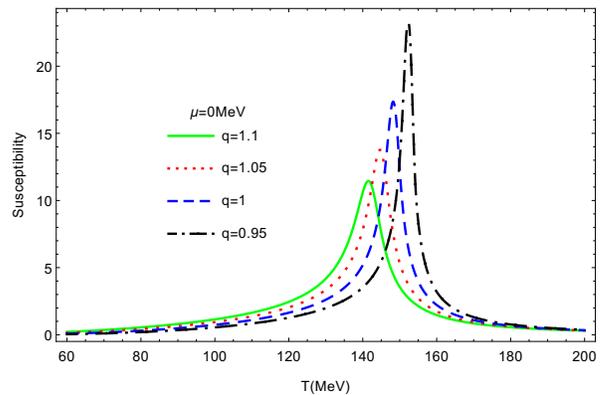}
\caption{The non-extensive effects on susceptibility 
$\chi =-\frac{\partial M}{\partial T}|_{\mu}$ with temperature $T$ at $\mu=0$
are shown for various $q$ as Fig.\ref{potential-M}.}
\label{x-T}
\end{figure}

Then the negative partial derivative of $M$ with respect to temperature $T$ holding chemical potential
$\mu$ constant, the susceptibility $\chi$,
is also investigated in this non-extensive linear sigma model, 
which describes the fluctuation of constituent (anti)quark mass.
From the results seen in Fig.\ref{x-T},
one can expect that, at the low density ($\mu=0$), the location of peak of susceptibility $\chi$,
as well as its own value, moves
to the lower values of temperature $T$ for larger $q$.
This indicates that with larger $q$, the critical temperature $T_c$
gets smaller, which supports it that the non-extensive
parameter $q$ describes the departure of system from the conditions of BG situation.

Here we add a few remarks to better understand the results.
Non-extensive dynamics develops the linear sigma model through
the (anti)quark number distribution functions. These functions are connected
with the thermal potential $\Omega_{\bar{\psi}\psi}(T, \mu, q)$ of Eq.(\ref{Therm-potential-q}),
which modifies the fluctuations of fermions.
The $q-$dependent chiral condensation can be obtained after
solving out the gap equation Eq.(\ref{gap-equ}).
From the upper panel of Fig.\ref{M-effect} we can see
its shape with respect to $T$ is strongly affected by the non-extensive
parameter $q$. 
More specifically, $q$ introduces differences of system itself from the usual
BG one 
which decrease the values of critical temperature $T_c$, seen in Fig.\ref{x-T}.

\begin{figure*}[htb]
\begin{center}
	\includegraphics[width = 0.75\linewidth]{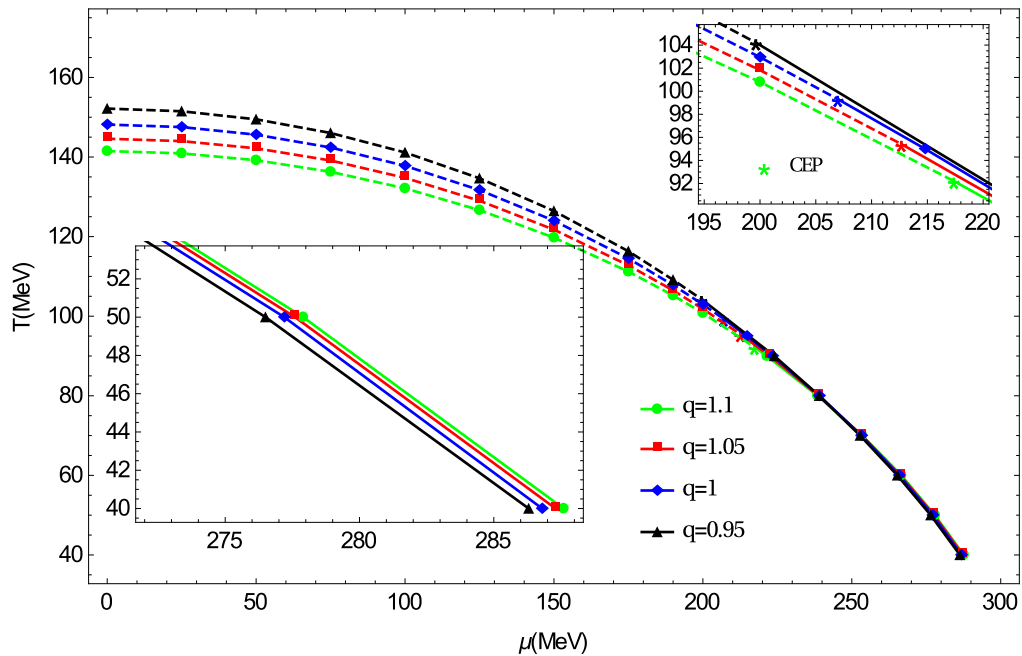}
\end{center}
\caption{Phase diagram of the $q-$linear sigma model in the $(T, \mu)$ plane for 
various $q$. The results are plotted for four different values of $q$ with
the vicinity of the $q-$dependent CEP and the low temperature part of the curves 
enlarged in the inset. For more details, the dashed line stands for crossover transition,
and solid one the first-order transition. CEPs are shown as star points.}
\label{T-mu}
\end{figure*}

In order to explore the chiral phase transition in the $q-$linear sigma model
more specifically, we also present the phase diagrams (seen in Fig.\ref{T-mu}) based
upon the analysis above. Easily seen that, indeed, at high temperature
and low density region it exhibits a crossover transition in the $(T-\mu)$
plane for different non-extensive parameters of $q$,
with smaller non-extensive parameter $q$ expanding 
the relative values of critical temperature and chemical potential. 
Meanwhile, a first-order phase
transition is shown at low temperature but high density region.
And all the critical lines correspondingly develop differently,
where larger $q$ increases the position of $T_c$ at the same $\mu_c$.

As for the critical endpoint (CEP), which locates between the two kinds of
phase transition, larger $q$ occurs to higher chemical potential
but lower temperature, which is also seen in the results of $q-$NJL model \cite{RW-09}. 
This is because systems from fewer
particles will encounter a larger value of $q$, whose phase transition
takes place with higher number density in turn.

\section{Summary and Outlook}\label{sec:sec4}

~~To summarize, we have calculated the non-extensive thermodynamics
of the chiral phase transition in the linear sigma model, to account for
the sensitivity of the mean field theory of the linear sigma model
to the departure from the usual BG statistics. By the $q-$version we have
obtained generalized relations of the grand canonical potential $\Omega$,
the chiral condensation $M$ and susceptibility $\chi$. Before that
we also analysis the values of non-extensive parameter $q$ and reasonably
consider the cases of $q=1.1, 1.05, 0.95$ as well as $q=1$.

Furthermore, we have investigated two scenarios,
$\mu \neq 0$ and $\mu=0$, respectively, which, as mentioned, correspond to
different physical situations: a first-order and a crossover transition.
For the studies of $\mu=0$, it is found to be in agreement with 
the results obtained in \cite{I-15}.
Besides, we discover it that different values of $q$ only influence
the quantities near the phase boundary. This also proves that
it is valuable and desirable to discuss the non-extensive
effects on the chiral phase transition.

As expected, the observed non-extensive effects of both the potential $\Omega$
and the mass $M$ lead to it that higher values of $q$ shift all to a earlier
state with other parameters fixed. In another word, the internal
divergence from the classical thermal equilibrium really impacts on the chiral phase
transition. This is more illuminated in the phase diagrams of $(T, \mu)$ plane correspondingly.
The CEP (see Fig.\ref{T-mu}) reveals a clear variation with different non-extensive
parameters of $q$, namely, holds higher chemical potential
but lower temperature with $q$ increasing, which agrees with 
Ref. \cite{RW-09}. 
As for the critical line in the diagram, as shown in Fig.\ref{T-mu},
$q-$effects derive different
trends of it on the first-order and crossover phase transitions,
whose physical mechanism needs us more attentions and investigations next.

Finally, It is worthy to mention that since CEP is still indistinct experimentally, our work
may provide a possible intensively study of locating the CEP in high-energy
physics \cite{R-09}. Meanwhile, by comparing the results with experimental data,
our researches could be of help to deeply understand the physical
explanation of the Tsallis non-extensive parameter $q$, which is also what we will
pay attention to in the future. 

\vspace{0.3cm}

\noindent {\bf Acknowledgments}

\vspace{3mm}

This research is partly supported by the Ministry of Science and Technology of China (MSTC) under the ``973" Project Nos. 2014CB845404, 2015CB856904(4), and by NSFC under Grant Nos. 11435004, 11322546, 11375070, 11521064.

\end{document}